\documentclass[usegraphicx]{mn2e}
\usepackage{psfig}
\catcode`\@=11
\def\gsim{\ifmmode{\mathrel{\mathpalette\@versim>}}
    \else{$\mathrel{\mathpalette\@versim>}$}\fi}
\def\lsim{\ifmmode{\mathrel{\mathpalette\@versim<}}
    \else{$\mathrel{\mathpalette\@versim<}$}\fi}
\def\@versim#1#2{\lower 2.9truept \vbox{\baselineskip 0pt \lineskip
    0.5truept \ialign{$\m@th#1\hfil##\hfil$\crcr#2\crcr\sim\crcr}}}
\catcode`\@=12

\newcommand{\beq}{\begin{equation}}
\newcommand{\eeq}{\end{equation}}
\newcommand{\az}{{a_0}}
\newcommand{\phiN}{\phi^{\rm N}}
\newcommand{\rhodm}{\rho_{\rm DM}}
\newcommand{\phidm}{\phi_{\rm DM}}
\newcommand{\phiT}{\phi_{\rm T}}
\newcommand{\gv}{{\bf g}}
\newcommand{\gvN}{{\bf g}^{\rm N}}
\newcommand{\vc}{v_{\rm c}}
\newcommand{\vcN}{v_{\rm cN}}
\newcommand{\sigmaR}{\sigma_R}
\newcommand{\sigmaz}{\sigma_z}
\newcommand{\sigmap}{\sigma_{\rm p}}

\def\ttv{\rm v}
\def\tvphi{\ttv_{\varphi}}
\def\tvphim{\overline{\ttv}_{\varphi}}
\def\tvphis{\tvphi^2}
\def\tvphism{\overline{\tvphis}}
\def\vcphi{u_{\varphi}}
\def\sigphi{\sigma_{\varphi}}
\def\sigphis{\sigphi^2}

\newcommand{\bey}{\begin{eqnarray}}
\newcommand{\eey}{\end{eqnarray}}

\newcommand{\kpc}{\, {\rm kpc} }

\newcommand{\kms}{\, {\rm km \, s}^{-1} }

\newcommand{\xv}{{\bf x}}

\newcommand{\Rz}{R_0}
\newcommand{\zz}{z_0}

\newcommand{\gNz}{g_z^{\rm N}}

\newcommand{\gzdisk}{g_{z,{\rm disk}}}
\newcommand{\gNzDM}{g^{\rm N}_{z,{\rm DM}}}

\def\los{{\it los }}
\def\xv{{\bf x}}

\def\nv{{\bf n}}
\def\nci{n_i}
\def\ncj{n_j}

\def\vv{{\bf v}}

\def\vci{v_i}

\def\vn{v_{\rm n}}
\def\vp{v_{\rm p}}

\def\Vp{V_{\rm p}}
\def\Vps{\Vp^2}
\def\sigij{\sigma_{ij}}

\def\sign{\sigma_{\rm n}}
\def\signs{\sign^2}
\def\sigob{\sigma_{\rm los}}

\def\sigobs{\sigob^2}

\def\Rtild{\tilde R}
\def\rtild{\tilde r}
\def\ztild{\tilde z}

\def\vphidb{\overline{{\rm v}_{\varphi}^2}}

\def\vn{u_{\rm n}}
\def\sign{\sigma_{\rm n}}

\def\vci{u_i}
\def\nv{{\bf n}}
\def\nci{n_i}
\def\ncj{n_j}
\def\sigij{\sigma_{ij}}
\def\siglos{\sigma_{\rm los}}

\def\xv{{\bf x}}

\def\Nr{N_r}
\def\NR{N_R}
\def\Nz{N_z}
\def\Nth{N_{\vartheta}}
\def\Nph{N_{\varphi}}
\newcommand{\Sv}{{\bf S}}

   \title[Vertical dynamics of disk galaxies in MOND]{Vertical
     dynamics of disk galaxies in MOND}

   \author[Nipoti et al.]
          {Carlo Nipoti$^1$, Pasquale Londrillo$^2$, Hong Sheng Zhao$^3$, and 
           Luca Ciotti$^1$
           \\ $^1$Astronomy Department, University of Bologna, 
                       via Ranzani 1, 40127 Bologna, Italy
           \\ $^2$INAF-Bologna Astronomical Observatory, I-40127 Bologna, 
              Italy
           \\ $^3$SUPA, University of St. Andrews, KY16 9SS, Fife, UK
           }

\date{Accepted 2007 April 3. Received 2007 March 28; in original form 2007 February 12.}
\pubyear{2007}

\begin{document} 
\maketitle

\begin{abstract} 

We investigate the possibility of discriminating between Modified
Newtonian Dynamics (MOND) and Newtonian gravity with dark matter, by
studying the vertical dynamics of disk galaxies. We consider models
with the same circular velocity in the equatorial plane (purely
baryonic disks in MOND and the same disks in Newtonian gravity
embedded in spherical dark matter haloes), and we construct their
intrinsic and projected kinematical fields by solving the Jeans
equations under the assumption of a two-integral distribution
function.  We found that the vertical velocity dispersion of deep-MOND
disks can be much larger than in the equivalent spherical Newtonian
models.  However, in the more realistic case of high-surface density
disks this effect is significantly reduced, casting doubts on the
possibility of discriminating between MOND and Newtonian gravity with
dark matter by using current observations.

\end{abstract}


\section{Introduction}
\label{secint}

The flatness of the rotation curves of disk galaxies in their external
regions has been the primary focus in the lasting on dark matter (DM)
and Modified Newtonian Dynamics (MOND).  As a consequence of the
formalisation of MOND in fundamental physics (Bekenstein~2004;
Zlosnik, Ferreira \& Starkman 2007), combined with some difficulties
of the cold dark matter scenario on galaxy scales (e.g. Binney~2004,
and references therein), the debate extended to other astrophysical
contexts.  For example, MOND models have been recently tested against
the Cosmic Background Radiation (Skordis et al.~2006, Skordis~2006),
gravitational lensing (Chen \& Zhao 2006; Zhao et al.~2006; Angus et
al.~2007; Takahashi \& Chiba~2007), star cluster and galaxy dynamics
(Zhao \& Tian 2006; Sanchez-Salcedo, Reyes-Iturbide \& Hernandez 2006;
Haghi, Rahvar \& Hasani-Zonooz~2006; Nipoti, Londrillo \& Ciotti~2007;
Scarpa et al.~2007; Tiret \& Combes~2007), and the solar system
(Bekenstein \& Magueijo~2006, Sanders~2006, Sereno \&
Jetzer~2006). Due to the surprising ability of MOND to reproduce the
kinematics of different systems (e.g., Milgrom~2002; Sanders \&
McGaugh~2002; Bekenstein~2006), it is of obvious interest to look for
other tests to discriminate between DM and MOND.

In this paper we explore the possibility of differentiating DM and
MOND by studying the {\it vertical dynamics} of disk galaxies. While
it is often taken for granted that---as far as disk kinematics is
concerned---the DM and MOND interpretations are nearly degenerate, in
fact this is not true.  An important issue is the choice of the
Newtonian system used for comparison with the MOND results. In the
present context, the proper comparison is between a baryonic disk in
MOND, and the same disk, in Newtonian gravity, immersed in a DM halo
(which for simplicity we assume spherically symmetric), such that the
circular velocity in the equatorial plane is the same as in the MOND
model\footnote{Other cases are known in which MOND and Newtonian
systems are equivalent with respect to some dynamical properties, but
different with respect to others (e.g. Ciotti \& Binney~2004).}. In
particular, we compare the kinematical fields of equivalent models
under the assumption of two-integral distribution function.  A work in
some respect complementary to ours, though with important differences,
was done by Milgrom~(2001), who studied, for a given baryonic disk,
the shape of the DM halo of the Newtonian model with exactly the same
dynamics (not only the same rotation curves) as the disk in MOND
gravity. Recently, a test of MOND based on the comparison of disk
circular velocity and vertical velocity dispersion has been proposed
also by Stubbs \& Garg (2005)\footnote{The presented test is affected
by the unjustified assumption that the vertical velocity dispersion
depends only on the local gravitational field.}.

The paper is organised as follows. In Section~\ref{secmet} we describe
the general method adopted, and we perform a preliminary investigation
of the problem. In Section~\ref{secres} we study in detail the
projected kinematical fields of Miyamoto-Nagai and thick exponential
disks in deep MOND regime, and we compare them with their equivalent
Newtonian models with DM, for different values of
flattening/thickness. We also address the problem of MOND dynamics in
Milky-Way like galaxy models. Our results are summarised in
Section~\ref{secdis}.

\section{The method}
\label{secmet}

In the present work we consider MOND in Bekenstein \& Milgrom's (1984)
formulation, in which the Poisson equation
\begin{equation}
\nabla^2\phiN=4\pi G\rho,
\label{eqPoisson}
\end{equation}
where $\phiN$ is the Newtonian gravitational potential generated by the
density distribution $\rho$, is substituted by the non-relativistic
field equation
\begin{equation}
\nabla\cdot\left[\mu\left({\Vert\nabla\phi\Vert\over\az}\right)
                 \nabla\phi\right] = 4\pi G \rho.
\label{eqMOND}
\end{equation}
In equation above $\Vert ...\Vert$ is the standard Euclidean norm,
$\phi$ is the MOND gravitational potential produced by $\rho$, and for
finite mass systems $\nabla\phi\to 0$ for $\Vert\xv\Vert\to\infty$.
The interpolating function $\mu(t)$ is not constrained by theory
except that it must run smoothly from $\mu(t)\sim t$ at $t\ll 1$ to
$\mu(t)\sim 1$ at $t\gg 1$, with a dividing acceleration scale $\az
\simeq 1.2 \times 10^{-10} {\rm m}\,{\rm s}^{-2}$.  In the so-called
`deep MOND regime' (hereafter dMOND), describing low-acceleration
systems ($\Vert\nabla\phi\Vert \ll\az$), $\mu(t)=t$ and
equation~(\ref{eqMOND}) reduces to
\begin{equation}
\nabla\cdot\left({\Vert\nabla\phi\Vert}\nabla\phi\right) = 4\pi G \az \rho.
\label{eqdMOND}
\end{equation}
As well known, equation~(\ref{eqMOND}) can be combined with
equation~(\ref{eqPoisson}) so that the MOND $\gv=-\nabla\phi$ and
Newtonian $\gvN=-\nabla \phiN$ gravitational fields are linked as
\begin{equation}
{\mu}(g/\az) \, \gv = \gvN +\Sv,
\label{eqmu}
\end{equation}
where $\Sv$ is a solenoidal field dependent on the specific $\rho$
considered. It can be proved that the equation above reduces to
Milgrom's (1983) empirical relation (i.e., $\Sv=0$) in case of
one-dimensional symmetries and in the special case of the razor-thin
Kuzmin~(1956) disk density distribution (Brada \& Milgrom~1995), but
in general one cannot impose $\Sv=0$. The presented results are
derived by solving the field equation~(\ref{eqMOND}) with the
numerical potential solver presented in Ciotti, Londrillo \&
Nipoti~(2006).

\subsection{Two-integral dynamics}

In the following we compare the dynamics of two disk galaxy models
with the same circular velocity curve in the equatorial plane
\begin{equation}
\vc^2(R)=R {\partial \phiT(R,0)\over \partial R},
\label{eqvc}
\end{equation} 
where $\phiT(R,z)$ is the total potential of the model, and
$(R,z,\varphi)$ are the standard cylindrical coordinates.  One model
is built in MOND for a baryonic disk galaxy described by the density
distribution $\rho(R,z)$.  The associated Newtonian model, which we
call {\it equivalent}, consists of the same baryonic disk plus a
spherical DM halo, whose radial density profile is determined by the
condition of matching the rotation curve obtained in the MOND model.
In practise, the density distribution of the DM halo of the equivalent
model is given by
\begin{equation}
\rhodm(r)={1\over{4\pi G r^2}}{d\over dr}\left[r\vc^2(r)-r\vcN^2(r)\right],
\label{eqrhodm}
\end{equation} 
where $\vc$ and $\vcN(R)$ are the disk circular velocity in MOND and
Newtonian gravity, respectively.  Note that the positivity of $\rhodm$
is not guaranteed for a generic disk density, so we always check the
positivity of the equivalent DM halo density distribution.

For a two-integral distribution function, i.e., $f=f(E,J_z)$, the
Jeans equations for the disk are
\begin{equation}
{\partial\rho\sigma^2\over \partial z}=-\rho {\partial\phiT\over\partial z},
\label{eqjeansz}
\end{equation}
\begin{equation}
{\partial\rho\sigma^2\over\partial R}+
{\rho(\sigma^2-\tvphism)\over R}=-
\rho {\partial\phiT\over\partial R},
\label{eqjeansR}
\end{equation}
where $\sigma^2\equiv \sigmaR^2=\sigmaz^2$,
$\tvphism=\vcphi^2+\sigphis$, $\vcphi=\tvphim$, and a bar over a
symbol indicates its phase-space mean over the velocity space (e.g., see
Binney \& Tremaine~1987, hereafter BT). Note that in a MOND model
$\phiT$ is the solution of equation~(\ref{eqMOND}) for the disk
density distribution, while in the equivalent Newtonian model
$\phiT=\phiN(R,z)+\phidm(r)$, with $\phidm$ determined from the
density profile in equation (\ref{eqrhodm}).

The velocity dispersion $\sigma^2(R,z)$ is obtained from integration
of equation (\ref{eqjeansz}) with boundary condition $\rho\sigma^2=0$
for $z\to\infty$, i.e.
\begin{equation} 
\rho\sigma^2=\int_z^\infty\rho {\partial\phiT\over\partial z'} dz'.
\label{eqsigma}
\end{equation}
In the equivalent model
\begin{equation} 
\rho\sigma^2=\int_z^\infty\rho {\partial\phiN\over\partial z'} dz' +
                          \int_z^\infty \rho 
                          {d\phidm\over dr}{z'\over r}  dz',
\label{eqsigmaequiv}
\end{equation}
where $r=\sqrt{R^2 +z'^2}$ and $d\phidm/dr =(\vc^2 -\vcN^2)/r$.

To split $\tvphism$ into streaming motion $\vcphi$ (that for
simplicity we assume nowhere negative), and azimuthal dispersion
$\sigphis$, we adopt the Satoh~(1980) $k$--decomposition
\begin{equation}
\vcphi^2 = k^2(\tvphism -\sigma^2),
\label{equphi}
\end{equation}
and
\begin{equation}
\sigphis =\sigma^2 + (1-k^2) (\tvphism - \sigma^2),
\label{eqsigphi}
\end{equation}
with $0\le k\le 1$; this procedure can be applied only when
$\vphidb-\sigma^2\geq 0$ everywhere.  For $k=0$ no ordered motions are
present, and the velocity dispersion tensor is maximally tangentially
anisotropic, while in the isotropic rotator ($k=1$) the galaxy
flattening is due to azimuthal streaming velocity.  In principle, by
allowing for $k=k(R,z)$, even more rotationally supported models can
be constructed, up to the maximum rotation case considered in Ciotti
\& Pellegrini~(1996), where $\sigphis=0$ everywhere.

%%%%%%%%%% FIG 1
\begin{figure}
\centerline{
\psfig{file=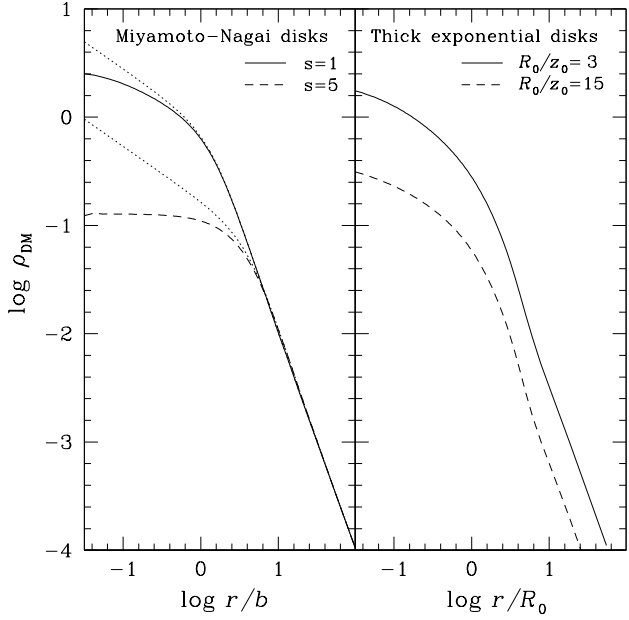,width=1.0\hsize}}
\caption{Density profile of spherical DM haloes of Newtonian models
equivalent to dMOND MN disks (left) and thick exponential disks
(right). The dotted lines in the left panel are given by the
distribution~(\ref{eqmnrhodm}), obtained in the $\Sv=0$
approximation. Density is in units of $(M/4\pi b^3) \sqrt{\az b^2/GM}$
(left) and $(M/4\pi R_0^2z_0) \sqrt{\az R_0^2/GM}$ (right).}
\label{figrho}
\end{figure}
%%%%%%%%%%%%%%%%%%%%%%%%%%

The explicit projection formulae for the kinematical fields of
axisymmetric two-integral systems viewed along a generic direction
$\nv$ of the line of sight ({\it los}) can be found elsewhere (e.g.,
see Lanzoni \& Ciotti~2003, Ciotti \& Bertin 2005, Riciputi et
al.~2005): here we just recall that at any given place in the galaxy
the \los component of the streaming velocity field is $\vn \equiv
\overline{\langle\vv ,\nv\rangle} =\vci\nci$, where ${\langle
  ,\rangle}$ is the standard inner product, the repeated index
convention is applied, and Cartesian coordinates are assumed.  The
analogous quantity associated with the velocity dispersion tensor is
$\signs\equiv \overline{\langle \vv-{\bf u}\rangle^2}=\sigij\nci\ncj$,
and the corresponding (mass--weighted) kinematical projected fields
are obtained by integration along the \los as
\begin{equation}
\Sigma\vp\equiv\int_{-\infty}^{\infty}\rho\vn dl,
\label{eqvp}
\end{equation}
\begin{equation}
\Sigma\Vps\equiv\int_{-\infty}^{\infty}\rho\vn^2 dl,
\label{eqVp}
\end{equation}
and
\begin{equation}
\Sigma\sigmap^2\equiv\int_{-\infty}^{\infty}\rho\signs dl,
\label{eqsigmap}
\end{equation}
where $\Sigma=\int_{-\infty}^{\infty}\rho\,dl$ is the disk density
projected along $\nv$.  In general $\sigmap$ is {\it not} the velocity
dispersion measured in observations: in the presence of a
non--zero projected velocity field $\vp$, this quantity is
\begin{equation}
\Sigma\sigobs\equiv\int_{-\infty}^{\infty}\rho\,\,
\overline{\left(\langle\vv ,\nv\rangle - \vp\right)^2} dl
=\Sigma\times\left(\sigmap^2+\Vps-\vp^2\right).  
\label{eqsiglos}
\end{equation}

For simplicity, in this paper we focus on the limit cases of disk
observed face-on and edge-on. The face-on projected velocity
dispersion is given by 
\begin{equation}
\Sigma\sigmap^2 = 2 \int_0^\infty \rho\sigma^2 dz
                = 2\int_0^\infty\rho {\partial\phiT\over\partial z}z dz,
\label{eqsigmapfo}
\end{equation}
where $\Sigma(R)=2 \int_0^{\infty}\rho(R,z) dz$ is the face-on disk
density.  In the edge-on case, assuming the \los directed along the
$x$ axis, the \los direction in the natural coordinate system is given
by\footnote{The \los vector points towards the observer, and so
{\it positive} velocities correspond to a {\it blue--shift}.} $\nv
=(1,0,0)$ and the projection plane is $(y,z)$, so that
\begin{equation}
\vn=-\vcphi\sin\varphi,
\end{equation}
and
\begin{equation}
\signs =\sigma^2 +(1-k^2)(\tvphism-\sigma^2)\sin^2\varphi,
\end{equation}
where $\sin\varphi= y/R=y/\sqrt{y^2+x^2}$.  Note that, after
projection, the $y$ coordinate can been identified with $R$.

\subsection{A preliminary analysis}
\label{secpre}

%%%%%%%%%% FIG 2
\begin{figure}
\centerline{
\psfig{file=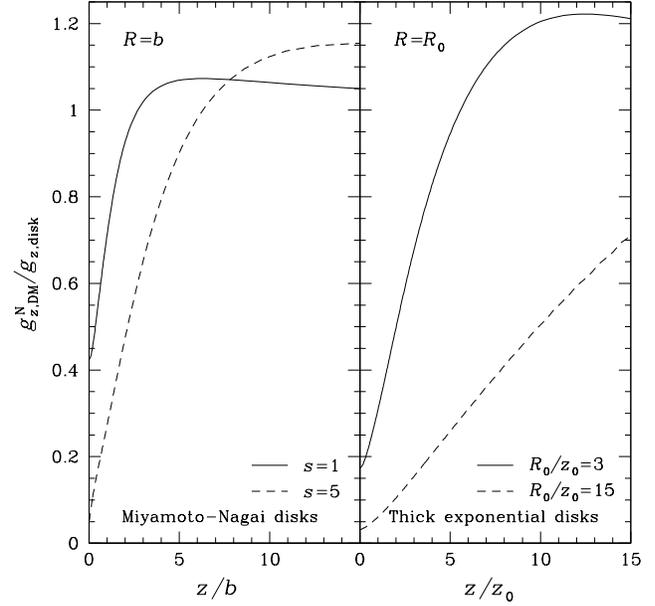,width=1.0\hsize}}
\caption{Ratio of the vertical accelerations of dMOND ($\gzdisk$) and
  equivalent Newtonian ($\gNzDM$) models as a function of the height
  $z$ above the disk plane, at fixed cylindrical radius.}
\label{figgz}
\end{figure}
%%%%%%%%%%%%%%%%%%%%%%%%%%

%%%%%%%%%% FIG 3
\begin{figure*}
\centerline{
\psfig{file=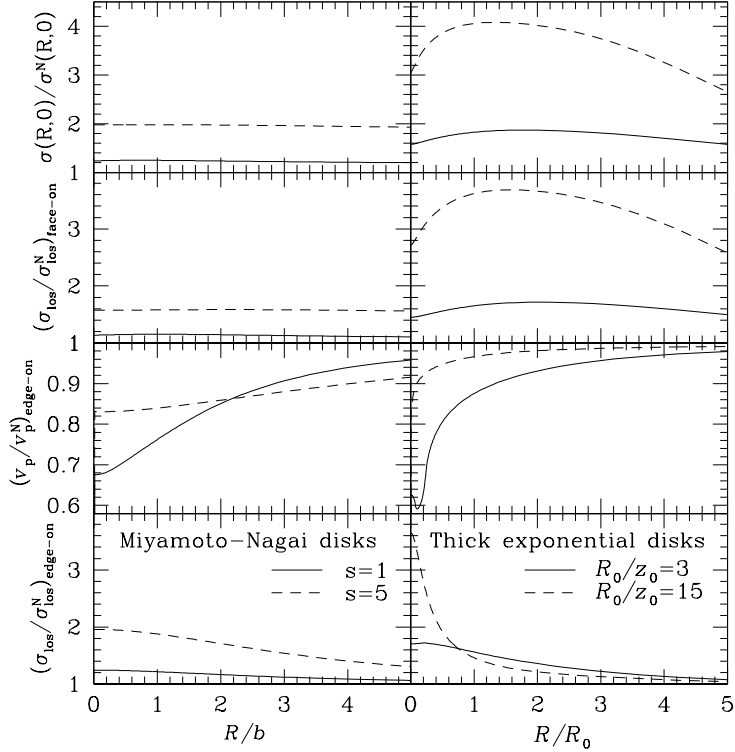,width=.6\hsize}}
\caption{Radial trend of velocity ratios between dMOND and equivalent
  Newtonian models (quantities with label N) in the fully isotropic
  case. From top to bottom: $z=0$ vertical velocity
  dispersion~(\ref{eqsigma}), face-on projected velocity
  dispersion~(\ref{eqsigmapfo}), $z=0$ edge-on projected streaming
  velocity~(\ref{eqvp}), and $z=0$ edge-on projected velocity
  dispersion (\ref{eqsiglos}).}
\label{figsig}
\end{figure*}
%%%%%%%%%%%%%%%%%%%%%%%%%%

As an introductory exercise we compare, as a function of disk
flattening, the vertical force field near the equatorial plane of a
Miyamoto \& Nagai (1975, hereafter MN) disk in dMOND regime and its
equivalent Newtonian model.  In order to carry out the calculations
analytically we assume $\Sv =0$, while in Section~\ref{secres} the
exact MOND gravitational field with the appropriate $\Sv$ field is
used.  The MN density distribution with scale length $b$ and
flattening parameter $s\equiv a/b$ is
\begin{equation} 
\rho(R,z)={M \over 4\pi b^3}
{s\Rtild^2+(s+3\zeta)(s+\zeta)^2\over[\Rtild^2+(s+\zeta)^2]^{5/2}\zeta^3},
\label{eqmnrho}
\end{equation} 
where $M$ is the total disk mass, $\zeta\equiv\sqrt{1+\ztild^2}$,
$\ztild\equiv z/b$, and $\Rtild\equiv R/b$.
The corresponding Newtonian potential is (BT)
\begin{equation}
\phi(R,z)=-{G M\over b}{1\over\sqrt{\Rtild^2 +(s+\zeta)^2}}.
\label{eqmnphin}
\end{equation}
For $a=0$ the MN density distribution is a Plummer~(1911) sphere with
scale radius $b$, while for $b=0$ one obtains the Kuzmin~(1956) disk.
The Newtonian disk circular velocity is given
by equations~(\ref{eqvc}) and~(\ref{eqmnphin}) as
\begin{equation}
\vcN^2(R)={G M\over b} {\Rtild^2\over [\Rtild^2+(s+1)^2]^{3/2}}.
\end{equation}
In the $\Sv=0$ approximation the dMOND circular velocity
is obtained from equation~(\ref{eqmu}) as
\begin{equation}
\vc^2(R)={\sqrt{G M \az} \Rtild^{3/2}\over [\Rtild^2+(s+1)^2]^{3/4}},
\end{equation}
and according to equation~(\ref{eqrhodm}), $\rhodm\propto
A(\rtild)-\sqrt{GM/\az b^2}B(\rtild)$, where $A$ and $B$ are two
dimensionless functions, and $\rtild\equiv r/b$.  In the dMOND limit
$\sqrt{GM/\az b^2} =0$ and so
\begin{equation}
\rhodm(r)\simeq  {M\over 4\pi b^3}\sqrt{\az b^2\over GM}
{2\rtild^2 + 5(s+1)^2\over 2\sqrt{\rtild}[\rtild^2+(s+1)^2]^{7/4}},
\label{eqmnrhodm}
\end{equation}
which is positive everywhere and has the expected $r^{-2}$ behaviour
for $r\to\infty$. Note that equation~(\ref{eqmnrhodm}) implies that in
the equivalent Newtonian model the MN disk is just a tracer of the DM
gravitational field.

The obtained formulae are useful because they allow to evaluate the
vertical gravitational field near the disk plane in the two models. In
the $\Sv=0$ approximation the dMOND vertical field is
\begin{equation} 
\gzdisk= \sqrt{\az\over ||\gvN||}\gNz,
\end{equation}
where $\gNz$ is the vertical Newtonian gravitational field of
the MN disk, and the vertical gravitational field of the equivalent
Newtonian DM halo is
\begin{equation} 
\gNzDM=-{\partial\phidm(r)\over\partial z}=-{z\vc^2(r)\over r^2}.
\end{equation}
Simple algebra then shows that for $R\to\infty$ and $z\to 0$, 
\begin{equation}
{\gNzDM \over \gzdisk}\sim {1\over 1+s}+\left[
{s\over 2 (1+s)^2} + {s(5+s)\over 4 (1+s)\Rtild^2}\right]\ztild^2,
\label{eqmngzrat}
\end{equation}
i.e., the vertical field is stronger in dMOND than in DM models, and
the discrepancy between the two cases is larger for flatter models
(i.e. for larger values of $s$). Thus, we expect $\sigma^2$ in the
disk to be different in the two models. In turn,
equations~(\ref{eqjeansR}) and (\ref{equphi}) imply that {\it also the
streaming velocity field in the disk equatorial plane is different in
the two models, even though the circular velocity is the same by
construction}.

\section{Results}
\label{secres}

We now obtain the kinematical fields (for simplicity in the fully
isotropic case) of a few disk galaxy models by computing the exact
MOND and Newtonian potentials with the numerical code presented in
Ciotti et al.~(2006). The potential solver is based on a spherical
grid of coordinates ($r$, $\vartheta$, $\varphi$), with
$\Nr\times\Nth\times\Nph$ points, uniformly spaced in ($\arctan{r}$,
$\vartheta$, $\varphi$).  To compute the intrinsic and projected
velocity dispersions, we interpolate the potential from the spherical
grid to a cylindrical grid with $\NR\times\Nz$ points, uniform in
($\arctan{R}$, $\arctan{z}$), on which we evaluate the integrals
(\ref{eqsigma}) and (\ref{eqvp}-\ref{eqsigmap}).  The numerical
integration routines are verified by deriving the intrinsic and
projected velocity dispersions of MN disks in Newtonian gravity, and
comparing the results with the analytical expressions reported in
Ciotti \& Pellegrini~(1996). For instance, with $\Nr=\Nth=\NR=\Nz=128$
the analytical results are reproduced with relative errors $\sim 0.1$
\%.

%%%%%%%%%% FIG 4
\begin{figure}
\centerline{
\psfig{file=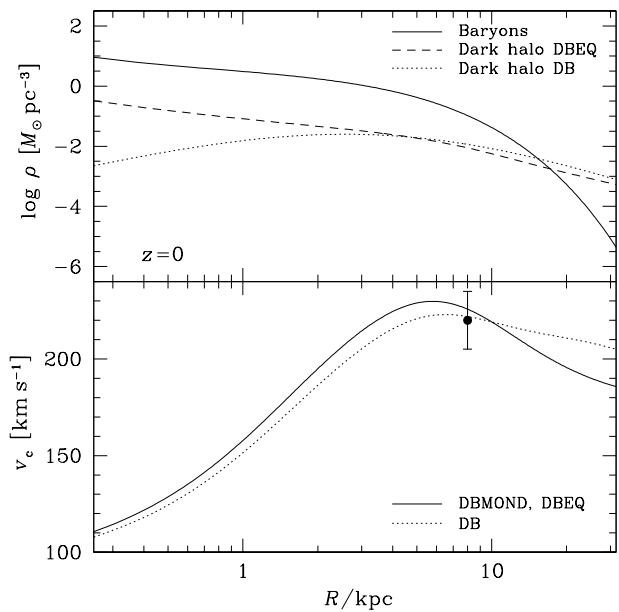,width=1.0\hsize}}
\caption{Density distribution in the equatorial plane for the baryonic
  and dark components of the considered Milky-Way like galaxy models
  (top), and the corresponding rotation curves (bottom).}
\label{figmwrho}
\end{figure}
%%%%%%%%%%%%%%%%%%%%%%%%

\subsection{Miyamoto-Nagai disks}
\label{secmnd}

We consider two MN galaxy models with flattening parameter $s=1$
(i.e., an almost spherical system), and $s=5$. Their exact dMOND
fields are computed by solving numerically equation~(\ref{eqdMOND}).
The density distributions of the spherical DM haloes of the
equivalent Newtonian models are then obtained from
equation~(\ref{eqrhodm}) with $\vcN=0$, as appropriate for the dMOND
case, in which the disk is just a tracer of the halo potential.  As
expected, the resulting halo profiles (Fig.~\ref{figrho}, left panel;
solid and dashed lines) decrease as $r^{-2}$ in the outer regions,
matching the analytical formula~(\ref{eqmnrhodm}), obtained in the
$\Sv=0$ approximation (dotted lines).  However, for $r\lsim b$ the
exact halo density profiles are considerably flatter than those
predicted by equation~(\ref{eqmnrhodm}), having inner logarithmic
slope $d\log \rhodm / d\log r \gsim -1/2$.  The difference (which is
larger in the flatter disk) indicates that the contribution of the
field $\Sv$ is non-negligible in the central regions, as also found by
Ciotti et al.~(2006) for triaxial and axisymmetric Hernquist~(1990)
models in dMOND regime. However, in the outer regions the agreement is
perfect.

In Fig.~\ref{figgz} (left panel), we plot the ratio $\gNzDM/\gzdisk$
as a function of $z$ at the representative radius $R=b$. Again, the
results nicely confirm the asymptotic estimate of
equation~(\ref{eqmngzrat}) when $z=0$: the effect is bigger in the
flatter disk, and just above the disk plane $\gzdisk$ is significantly
stronger than $\gNzDM$. We also note that for $z\gg b$ the vertical
Newtonian field exceeds the dMOND one, though slightly.  As a
consequence, also the intrinsic and projected velocity dispersions are
different in the equivalent Newtonian and dMOND MN disks, while the
circular velocities in the equatorial plane are the same by
construction. The midplane vertical velocity dispersion $\sigma(R,0)$
and the projected face-on velocity dispersion $\siglos$ are larger in
the dMOND systems than in the equivalent Newtonian systems at all
radii $R<5b$, with the larger discrepancies (up to a factor of 2) in
the flatter disk (Fig.~\ref{figsig}, left column).  The situation is
more complicate in the edge-on projection: the projected streaming
velocity $\vp$ in the Newtonian models are larger than in the dMOND
cases, while the edge-on projected velocity dispersion $\siglos$ in
the galactic plane is higher in dMOND than in Newtonian models.

\subsection{Thick exponential disks}
\label{secexp}

We now investigate the more realistic thick exponential disk
\begin{equation} 
\rho(R,z)={M\over 4\pi\Rz^2\zz}\exp\left(-{R\over\Rz}\right)
            {\rm sech}^2\left({z\over\zz}\right),  
\label{eqexpd}
\end{equation} 
where $M$ is the total disk mass, and $\Rz$ and $\zz$ are
characteristic scale-lengths. As for the MN case, we consider here two
dMOND models (and so in the equivalent Newtonian model the disk does
not contribute to the potential): a thicker disk with $\Rz=3\zz$, and
a thinner disk with $\Rz=15\zz$. 

%%%%%%%%%% FIG 5
\begin{figure}
\centerline{
\psfig{file=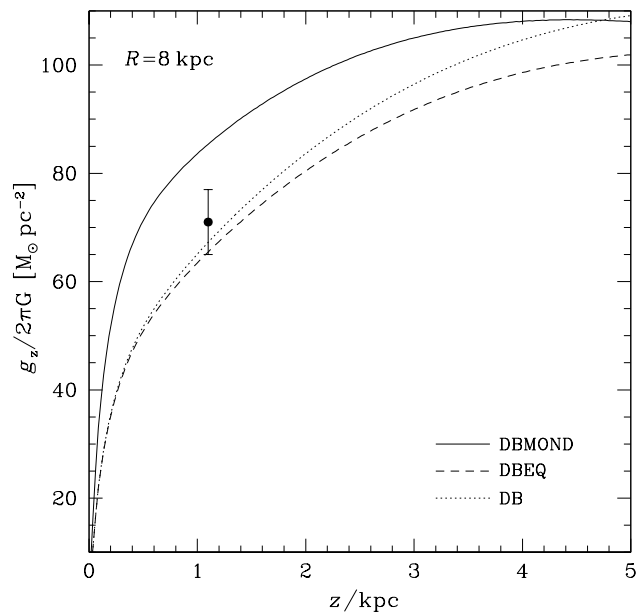,width=1.0\hsize}}
\caption{Vertical force as a function of $z$, at $R=8\kpc$, for the
  considered Milky-Way like galaxy models.}
\label{figmwgz}
\end{figure}
%%%%%%%%%%%%%%%%%%%%%%%%%%

As in MN disks, the spherical DM haloes density profiles of the
equivalent Newtonian models are rather flat in the central regions and
$\propto r^{-2}$ at large radii (Fig.~\ref{figrho}, right
panel). Interestingly, these DM haloes ``predicted'' by MOND for
low-surface brightness disks are not characterised by the steep
central cusps expected in the context of cold dark matter.  The dMOND
vertical force near the disk is significantly stronger than in the
equivalent Newtonian models (Fig.~\ref{figgz}, right panel): at
$R=\Rz$ in the thinner model the dMOND vertical force ratio is $\sim
10$ just above the plane.  This difference reflects in the velocity
fields (Fig.~\ref{figsig}, right panels), which behave qualitatively
as the MN models, but the discrepancies are larger: for example, the
ratio of intrinsic velocity dispersion is as high as $\sim 4$ in the
thinner model.

%%%%%%%%%% FIG 6
\begin{figure*}
\centerline{
\psfig{file=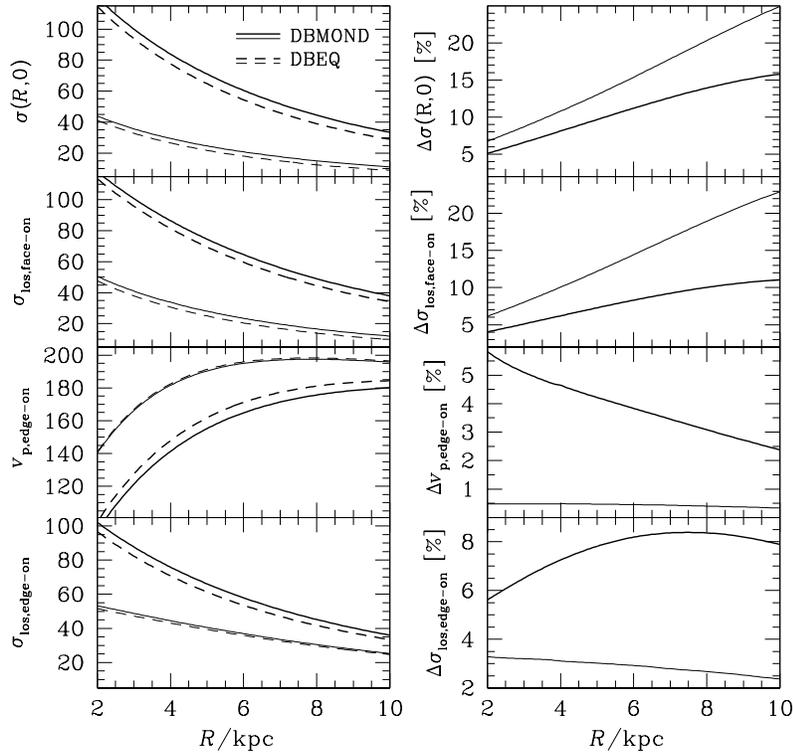,width=.6\hsize}}
\caption{Left (from top to bottom): $z=0$ vertical velocity
  dispersion, face-on projected velocity dispersion, $z=0$ edge-on
  projected streaming velocity and $z=0$ edge-on projected velocity
  dispersion for model DBMOND (solid lines) and its equivalent
  Newtonian model DBEQ (dashed lines), in units of $\kms$.  Right:
  percentage absolute differences between the corresponding MOND and
  Newtonian velocity fields in the left panels. Thin and thick lines
  refer to thin-disk and thick-disk stars, respectively.}
\label{figmwsig}
\end{figure*}
%%%%%%%%%%%%%%%%%%%%%%%%%%

%%%%%%%%%% FIG 7

\begin{figure*}
\centering
\includegraphics[width=0.3\textwidth,viewport=5 5 370 370,clip]{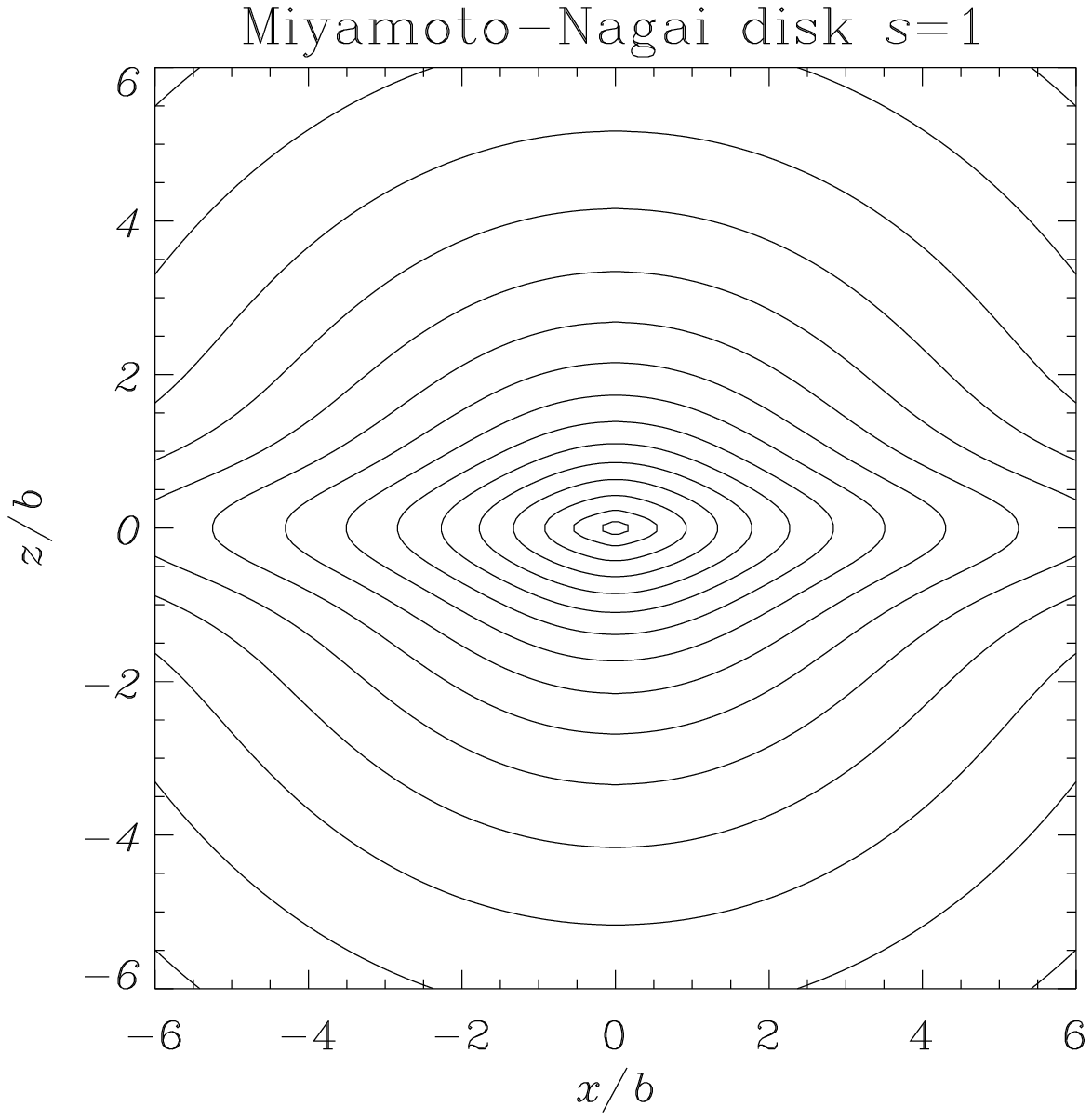}
%\hfill
\includegraphics[width=0.3\textwidth,viewport=5 5 370 370,clip]{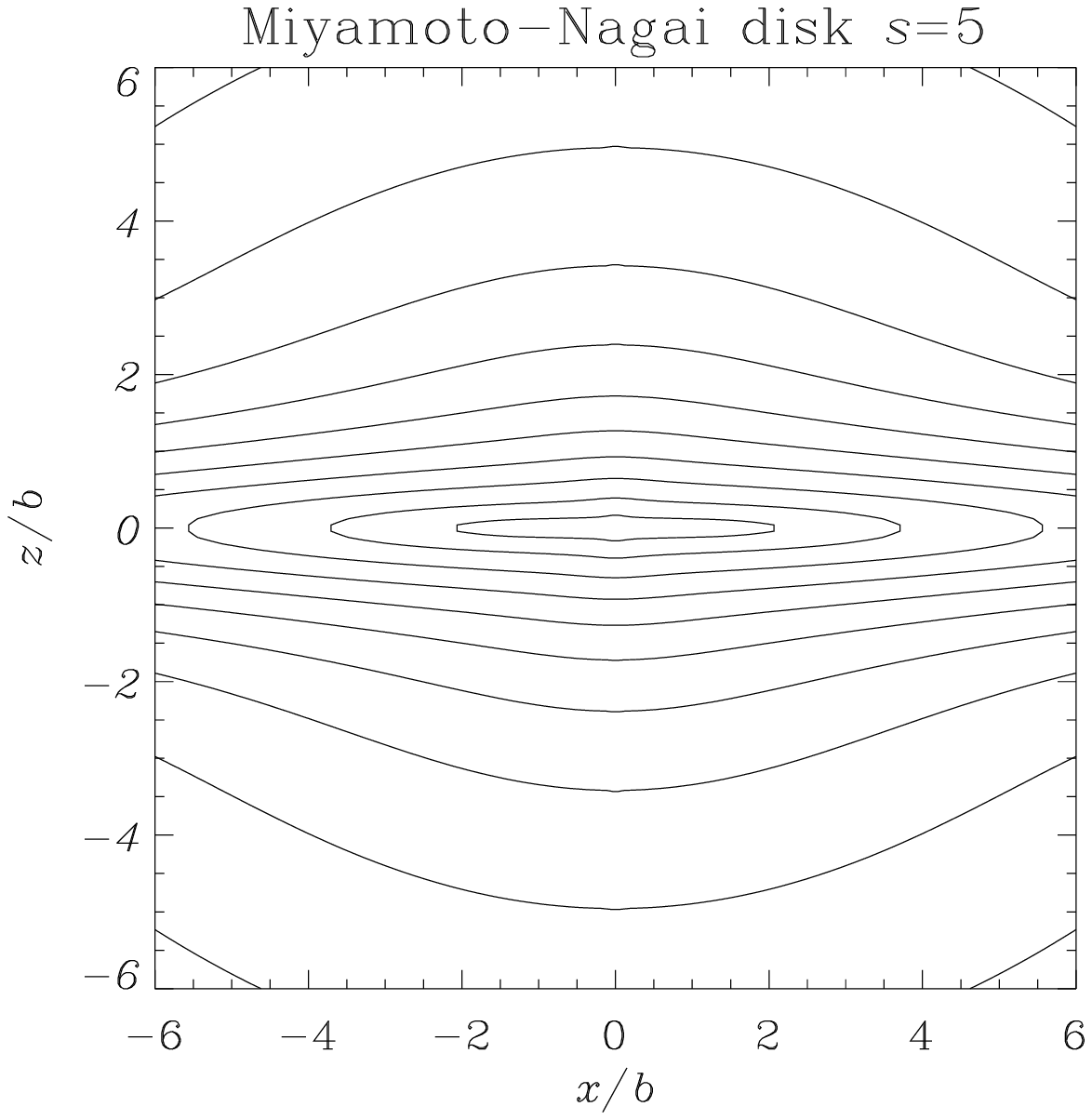}\\
\includegraphics[width=0.3\textwidth,viewport=5 5 370 370,clip]{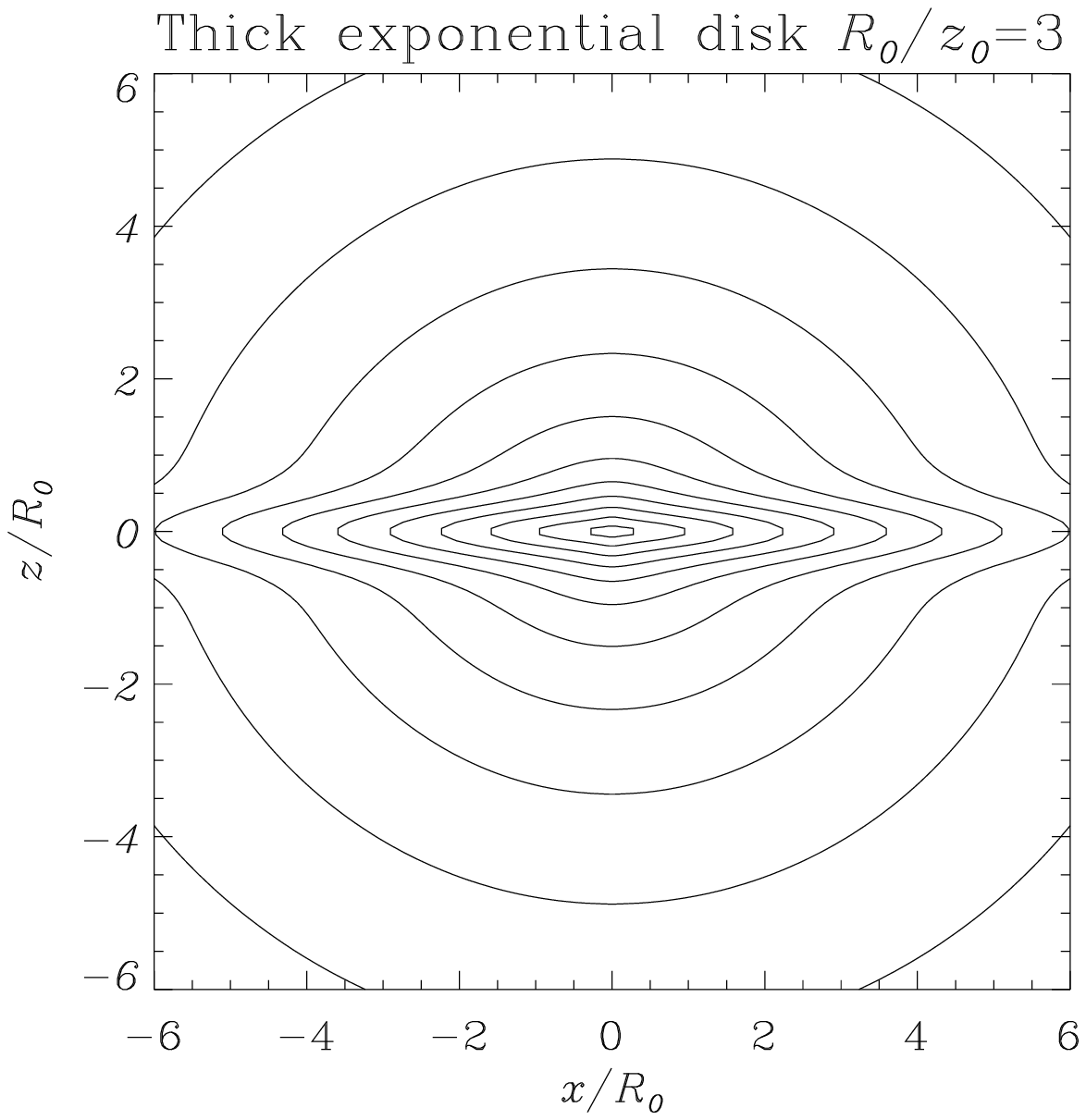}
%\hfill
\includegraphics[width=0.3\textwidth,viewport=5 5 370 370,clip]{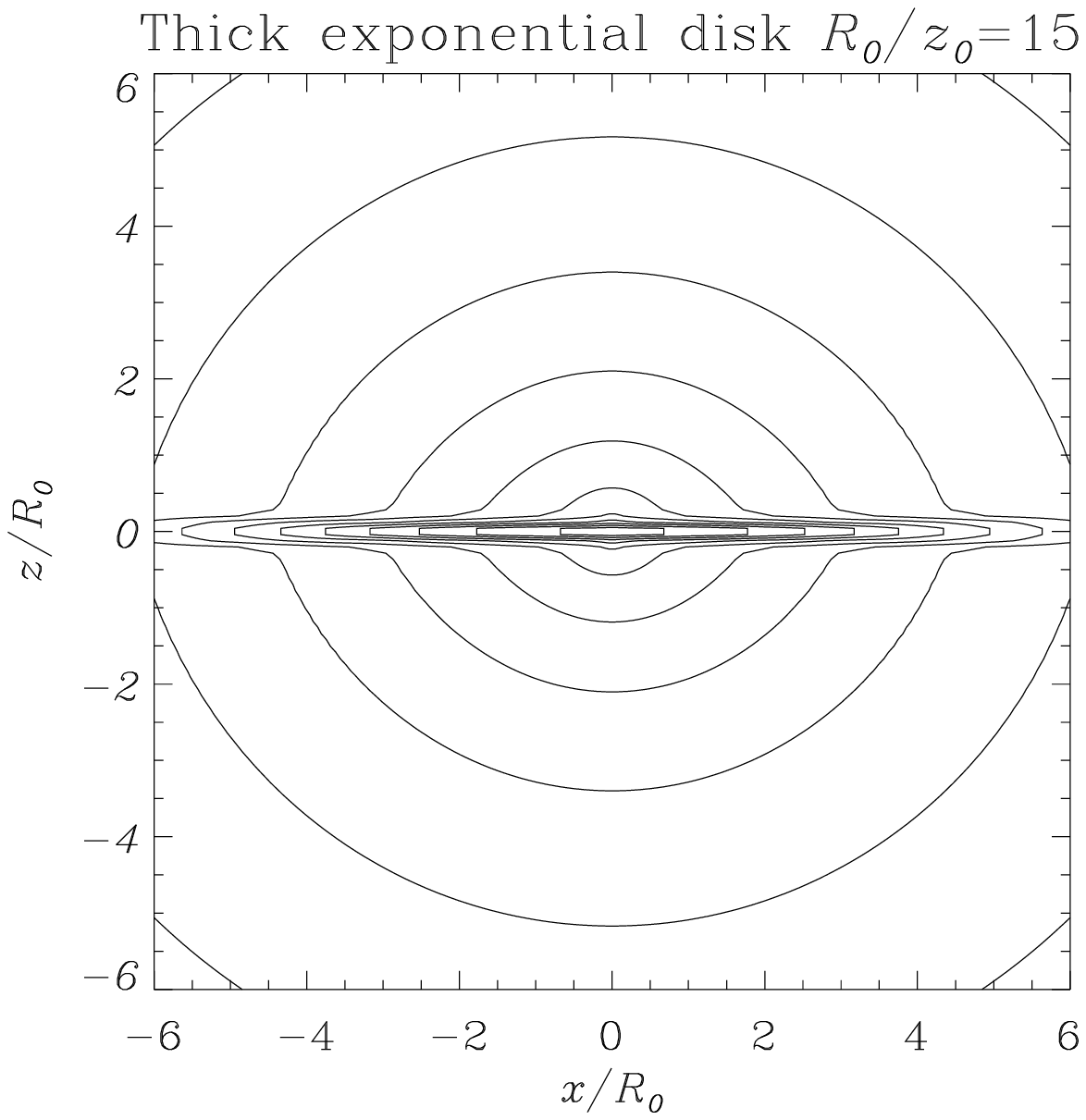}
\\
\caption{Isodensity contours in the meridional plane of the fully
  equivalent DM haloes associated with dMOND MN (top panels) and thick
  exponential disks (bottom panels). Density increases towards the
  equatorial plane.}\label{figflathalo}
\end{figure*}
%%%%%%%%%%%%%%%%%%%%%%%%%%

\subsection{Milky-Way like galaxies}

The previous analysis restricted to models in the dMOND regime, in
order to estimate the largest possible differences in the kinematical
fields of equivalent disk galaxies. We now focus on a Milky-Way like
galaxy model, with the purpose of quantifying the MOND kinematical
effects in realistic high surface-brightness galaxies.  We will {\it
not} enter the challenging problem of finding the best-fitting MOND
model of the Milky Way (which is beyond the aim of the present paper;
see Famaey \& Binney~2005); instead, we apply our method to one of the
currently available mass models of the Milky Way that have been
derived in the context of DM.  In particular, we consider
Model~1 of Dehnen \& Binney (1998; hereafter model DB), which consists
of a truncated oblate spheroidal power-law bulge, a stellar disk made
of a thin and a thick component, a gaseous exponential disk with a
central hole, and finally an oblate spheroidal DM distribution.  Model
DB represents a good Newtonian dynamical model of the solar
neighbourhood and the Galaxy, and can be classified as a disk
dominated model, because in Newtonian gravity the disk provides the
main contribution to the rotation curve in the central regions.

As we solve now equation~(\ref{eqMOND}), we need to specify the
interpolating function $\mu$.  Following Famaey \& Binney~(2005), we
fix $\az \simeq 1.2 \times 10^{-10} {\rm m}\,{\rm s}^{-2}$ and
\begin{equation}
\mu(t)={t \over 1+t}.  
\label{eqmusimple}
\end{equation}
This function gives a better fit to the terminal velocity curve of the
Milky Way than the commonly adopted function $\mu(t)=t/\sqrt{1+t^2}$
and fits extremely well the rotation curve of NGC3198, also being
consistent with Bekenstein's (2004) TeVeS theory (Zhao \& Famaey~2006;
see also Famaey et al.~2007).  We consider two models with the same
baryonic distribution as model DB: model DBMOND (a purely baryonic
MOND model), and model DBEQ (the equivalent Newtonian model with
spherical DM halo). Clearly, the DM halo density profile in model DBEQ
is not identical to that of model DB: Fig.~\ref{figmwrho} (top panel)
shows that in the equatorial plane the two distributions are very
similar in the radial range $R\sim5-15\kpc$, while the halo profile of
DBEQ is steeper than that of DB at small radii.  As a consequence, the
rotation curve of models DBMOND and DBEQ (which is the same by
construction) differs from the rotation curve of DB, being
systematically higher at small radii and lower at large radii
(Fig.~\ref{figmwrho}, bottom). However, the differences in the radial
range $0-15\kpc$ are within $\sim5\%$, and $\vc(8\kpc)=226\kms$ in
model DBMOND (and DBEQ), consistent with the observational constraint
at the solar neighbourhood $220\pm15\kms$ (solid symbol and vertical
bar in the diagram; e.g., BT).

In contrast with the dMOND cases discussed in Sections~\ref{secmnd}
and \ref{secexp}, now the contribution of the baryonic component to
the gravitational field is important also in the Newtonian model, and
this results in smaller differences between the MOND and Newtonian
equivalent models. For instance, considering the vertical force as a
function of $z$ at $R=8\kpc$ (Fig.~\ref{figmwgz}), we find that the
MOND force is stronger than the Newtonian one, with a difference of
$\sim30\%$ at $z\sim$1~kpc. The solid dot marks an observational
estimate (and associated uncertainty) of the vertical force 1.1~kpc
above the plane at the solar radius (Kuijken \& Gilmore 1989,
1991). Remarkably, the difference between the vertical forces at
$z=1.1\kpc$ predicted by the MOND and Newtonian models is larger than
the observational error. So, taken at face value, this result would
favour DM rather than MOND models, though we stress again that we did
not attempt to build the best-fit MOND model of the Milky Way, and we
cannot exclude the possibility of finding a MOND model of the Milky
Way satisfying this and other observational constraints (see also
Famaey \& Binney~2005). In any case, this result shows that in
principle it could be possible to try to falsify MOND by using the
observational constraints on the vertical force above the plane.

Following the treatment of Sections~\ref{secmnd} and \ref{secexp} we
computed the kinematical fields of models DBMOND and DBEQ, again under
the same assumptions of two-integral distribution
function\footnote{Because of this assumption our results on the
velocity fields cannot be intended, strictly speaking, to represent
the specific case of the Milky Way, whose distribution function is
{\it not} two-integral, being $\sigmaR^2>\sigmaz^2$ in the solar
neighbourhood.}  and full isotropy. For both models we computed the
intrinsic and projected velocity fields of the thin-disk (thin curves)
and thick-disk (thick curves) stars (Fig.~\ref{figmwsig}, left
panels), finding that the discrepancy between the two models is up to
20-25\% for intrinsic and face-on velocity dispersion, while just few
per cent in case of edge-on projection (Fig.~\ref{figmwsig}, right
panels).  These results show that the differences in the vertical
kinematical fields in a realistic high-surface density galaxy are
quite small, and it is not obvious that one can discriminate between
MOND and Newtonian gravity (plus DM) by using the information
currently derived by the observations.

\section{Discussion and conclusions}
\label{secdis}

In this paper we quantified the differences in the vertical force and
kinematical fields between MOND disk galaxy models and equivalent
(i.e., having identical circular velocity in the equatorial plane)
Newtonian models with spherical DM haloes. We showed that, in
principle, MOND and Newtonian gravity with DM can be differentiated
with accurate measurements of the vertical force near the disk and of
the projected kinematical fields. In particular, MOND models have
stronger vertical force near the plane and higher vertical velocity
dispersion than equivalent Newtonian models with DM. Our results are
not inconsistent with the finding of Read \& Moore~(2005) that
satellite orbits in DM and MOND behave very similarly, because far
from the disk plane the force fields predicted by the two theories are
very similar.

From the observational point of view, the robust discrimination
between MOND and Newtonian gravity with DM is still challenging due to
several factors: (i) the strongest effects are expected in
low-acceleration (low-surface-brightness) systems, in which
measurements of the velocity fields are difficult; (ii) there are
often several stellar populations with different scale-lengths and
heights, and the mass-to-light of the stellar component is tunable;
(iii) most disks have gaseous components whose density has non-smooth
features which could make differences in any local volume of kpc
scale; (iv) the models should be compared to data with noise so our
comparisons of MOND and DM models of exactly same circular velocity
curves remains a theoretical exercise; (v) DM haloes are not
necessarily spherically symmetric. 

With regard to point (v), we briefly discuss how our results would
change by relaxing the assumption of an equivalent {\it spherical} DM
halo.  Following Milgrom~(2001), we start by constructing, for a given
disk distribution $\rho$, the ``fully equivalent'' DM density
distribution as $\rhodm=\nabla^2\phi /4\pi G-\rho$, where $\phi$ is
the MOND potential; as well known, $\rhodm$ can be negative in some
region of space (Milgrom~1986; Ciotti et al.~2006). By construction,
the fully equivalent Newtonian model has gravitational field identical
to the MOND disk field in all the space, not just the same midplane
circular velocity (as the equivalent model with spherical halo).
Interestingly, we found that the density distributions $\rhodm$ of the
fully equivalent DM haloes of the dMOND models presented in
Sections~\ref{secmnd} and \ref{secexp} are nowhere negative (in these
cases $\rhodm=\nabla^2\phi /4\pi G$, because the baryonic disk is just
a tracer in dMOND-equivalent models).  Figure~\ref{figflathalo} shows
that $\rhodm$ presents a significant disk structure (more pronounced
in the case of more flattened baryonic disks). This result clearly
indicates that it is not possible to obtain a Newtonian model with
spheroidal DM distribution {\it strictly} equivalent to a MOND disk
model (see also Milgrom~2001).  However, we found that the discrepancy
in the vertical dynamics in the galactic plane between MOND and DM
models can be significantly reduced considering spheroidal DM haloes:
for instance, a dMOND MN disk with $s=5$ and an oblate spheroidal halo
with axis ratio $\sim 0.25-0.3$ (and the same asymptotic circular
velocity as the baryonic dMOND disk) would produce very similar
vertical velocity dispersion and rotational velocity in the galactic
plane (excluding the very central regions).

For all these reasons, it appears difficult to differentiate between
MOND and DM using the currently available observational measures of
stellar kinematics of external disk galaxies (e.g., Bottema~1993;
Kregel \& van der Kruit~2005, and references therein).  The most
promising application of our method seems to be the case of the Milky
Way: in particular, for the specific baryonic mass distribution
considered in this paper the difference between the MOND and DM
vertical forces above the plane is larger than the uncertainty on the
observational estimates.  The estimates of both the baryonic mass
distribution and the vertical force above the plane are expected to be
greatly improved in the near future thanks to data from the GAIA
mission.

\section*{Acknowledgments}
We thank Renzo Sancisi for helpful discussions, and the Referee,
Robert Sanders, for his useful suggestions. L.C. and P.L. were
partially supported by a MIUR grant Cofin 2004.


\begin{thebibliography}{99}

\bibitem{}Angus G.W., Shan H.Y., Zhao H.S., Famaey B.~2007, ApJ, 654, L13
\bibitem{}Bekenstein J., 2004, Phys. Rev. D, 70, 3509 
\bibitem{}Bekenstein J., 2006, Contemporary Physics, 47, 387 
\bibitem{}Bekenstein J., Magueijo J., 2006,  Phys. Rev. D, 73, 103513
\bibitem{}Bekenstein J., Milgrom M. 1984, ApJ, 286, 7 
\bibitem{}Binney, J. 2004, in Ryder, S.D., Pisano, D.J., Walker, M.A.,
  Freeman, K.C., eds, IAU Symp. 220, Dark Matter in Galaxies. Astron. Soc. Pac., San Francisco, p. 3
\bibitem{}Binney J., Tremaine S., 1987, Galactic Dynamics, Princeton University Press, Princeton (BT) 
\bibitem{}Bottema R., 1993, A\&A, 275, 16
\bibitem{}Brada R., Milgrom M. 1995, MNRAS, 276, 453
\bibitem{}Chen D.M., Zhao H.S., 2006, ApJ, 650, L9 
\bibitem{}Ciotti L., Bertin G., 2005, A\&A, 437, 419
\bibitem{}Ciotti L., Binney J., 2004, MNRAS, 351, 285
\bibitem{}Ciotti L., Londrillo P., Nipoti C.,   2006, ApJ, 640,  741
\bibitem{}Ciotti L., Pellegrini S., 1996, MNRAS, 279, 240
\bibitem{}Famaey B., Binney J., 2005, MNRAS, 363, 603 
\bibitem{}Famaey B., Gentile G., Bruneton J.P., Zhao H.S., 2007, Phys. Rev. D, 75, 063002 
\bibitem{}Haghi H., Rahvar S., Hasani-Zonooz A.,  2006, ApJ, 652, 354
\bibitem{}Hernquist L., 1990, ApJ, 356, 359
\bibitem{}Kregel M., van der Kruit P.C., 2005, MNRAS, 358, 481
\bibitem{}Kuijken K., Gilmore G., 1989, MNRAS, 239, 651
\bibitem{}Kuijken K., Gilmore G., 1991, ApJ, 367, L9 
\bibitem{}Kuzmin G., 1956 Astron. Zh, 33, 27
\bibitem{}Lanzoni B., Ciotti, L., 2003, A\&A, 404, 819
\bibitem{}Milgrom M., 1983, ApJ, 270, 365 
\bibitem{}Milgrom M., 1986, ApJ, 306, 9 
\bibitem{}Milgrom M., 2001, MNRAS, 326, 1261
\bibitem{}Milgrom M., 2002, New. Astron. Rev., 46, 741 
\bibitem{}Miyamoto M., Nagai R., 1975, PASJ, 27, 533 (MN)
\bibitem{}Nipoti C., Londrillo P., Ciotti L., 2007, ApJ, 660, 256
\bibitem{}Plummer H.C., 1911, MNRAS, 71, 460 
\bibitem{}Read J.I., Moore B.,  2005, MNRAS, 361, 971
\bibitem{}Riciputi A., Lanzoni B., Bonoli S., Ciotti, L., 2005, A\&A, 443, 133
\bibitem{}Sanders R.H, 2006, MNRAS, 370, 1519
\bibitem{}Sanders R., McGaugh S., 2002, ARA\&A, 40, 263,
\bibitem{}S\'anchez-Salcedo  F.J., Reyes-Iturbide  J., Hernandez X., 2006, MNRAS, 370, 1829
\bibitem{}Satoh C., 1980, PASJ, 32, 41
\bibitem{}Sereno M., Jetzer Ph., 2006, MNRAS, 371, 626
\bibitem{}Scarpa R., Marconi G., Gilmozzi R., Carraro G., 2007, A\&A, 462, L9
\bibitem{}Skordis C., Mota D.F., Ferreira P.G., Boehm C., 2006, Phys. Rev. Lett., 96, 011301
\bibitem{}Skordis C., 2006, Phys. Rev. D, 74, 103513
\bibitem{}Stubbs C.W., Garg A., 2005, preprint (arXiv:astro-ph/0512067v1)
\bibitem{}Takahashi R., Chiba T., 2007, preprint (arXiv:astro-ph/0701365v2)
\bibitem{}Tiret O., Combes F., 2007, A\&A, 464, 517
\bibitem{}Zhao H.S., Bacon D., Taylor A.N., Horne K.D., 2006,  MNRAS, 368, 171
\bibitem{}Zhao H.S., Famaey B.,  2006, ApJ, 638, L9
\bibitem{}Zhao H.S., Tian L.,  2006, A\&A, 450, 1005
\bibitem{}Zlosnik T.G., Ferreira P.G., Starkman G.D., 2007, Phys. Rev. D, 75, 044017
\end{thebibliography}
\end{document}